\documentclass [11pt] {article}

\newcommand{\bye}{\end{document}}

\newcommand{\be}{\begin{equation}}
\newcommand{\ee}{\end{equation}}
\newcommand{\beq}{\begin{eqnarray}}
\newcommand{\eeq}{\end{eqnarray}}

\newcommand{\llra}{\longleftrightarrow}

\newcommand{\lra}{\longrightarrow}

\newcommand{\muk}{\mu^+ K^0}

\newcommand{\nmk}{{\bar\nu}_\mu K^+}

\def\NPB#1#2#3{Nucl. Phys. {\bf B} {\bf#1} (19#2) #3}
\def\PLB#1#2#3{Phys. Lett. {\bf B} {\bf#1} (19#2) #3}
\def\PRD#1#2#3{Phys. Rev. {\bf D} {\bf#1} (19#2) #3}
\def\PRL#1#2#3{Phys. Rev. Lett. {\bf#1} (19#2) #3}
\def\PRT#1#2#3{Phys. Rep. {\bf#1} C (19#2) #3}

\begin{document}
\begin{titlepage}
\thispagestyle{empty}
\parindent 0pt
\begin{large}
\title{Large Right Handed Rotations, Neutrino Oscillations\\ and Proton
  Decay\footnote{Talk at the TAUP99 Paris, 6-10 Sept. 1999. To be
  published in JHEP.}}
\end{large}
\author{Yoav
 Achiman\footnote{e-mail:achiman@theorie.physik.uni-wuppertal.de}\quad
\quad and \quad Carsten Merten\footnote{e-mail:merten@theorie.physik.uni-wuppertal.de}\\
[0.5cm]
        Department of Physics\\
        University of Wuppertal\\
        Gau\ss{}str.~20, D--42097 Wuppertal \\
        Germany\\[1.5cm]}

\date{October 1999}

\maketitle
\setlength{\unitlength}{1cm}
\begin{picture}(5,1)(-12.5,-12)
\put(0,0){WUB 99-26}
\end{picture}
\parindent0cm

\begin{abstract}
Right Handed (RH) rotations are not observable in the standard model (SM).
The freedom is used to give the phenomenologically correct mass matrices
all kind of different forms and this is one of the reasons for the 
proliferation of models for the fermionic masses.\\
The SM must be however extended and  in most extensions, RH 
currents appear at higher energy scales. At those energies the RH
rotations are not 
irrelevant any more, they can affect neutrino oscillations, proton
decay, baryon assymetry, R-parity violating interactions etc.\\
We study possible implications of large RH mixing in GUTs. 
Those are interesting not only because large mixing induce large effects. They
are intimately related to large lepton mixing in GUTs, via a relation
between LH mixing of the leptons and the RH ones of the d-quarks
,(``$d - \ell$ duality'') and can change considerably the branching
ratios of proton decay.
 Observation of proton decay 
channels as well as neutrino oscillations will teach us about RH rotations
and will reduce, therefore, considerably the freedom in the fermionic mass 
matrices. Some interesting examples are studied in detail. In particular
a new $E_6$ model which realizes naturally a $d$ - $\ell$ duality by mixing
with exotic $E_6$ fermions.   
\end{abstract}
\thispagestyle{empty}

\end{titlepage}
\clearpage
\setcounter{page}{1}

Superkamiokande~\cite{sk} confirmed in this conference once again the
observation that large 
neutrino mixing is responsible for the atmospheric neutrino anomaly. This leads
to a puzzle: how can leptons have large left-handed (LH) mixing in clear 
contradiction with the small LH mixing of the quarks.  I am emphasizing  ``LH
 mixing''  because the fermions acquire also right-handed (RH) mixing. Those
are unobservable in the standard model (SM) but may play an important role in
its extensions~\cite{cor}~\cite{akb}. The aim of this talk is to emphasize
the importance of RH rotations and the fact that they may be the
solution of the 
mixing puzzle. In particular in terms of the {\em duality} observed in
 certain GUTs between d-quarks and charged 
leptons~\cite{5}~\cite{10}~\cite{bk}:\\
\centerline{\em RH rotations of d-quarks $\llra$\ LH rotations of
  charged leptons  } 
\centerline{\em LH rotations of d-quarks $\llra$\ RH rotations of
 charged leptons.}  
This means that the observed large LH mixing of the leptons corresponds not to
 the LH mixing of the quarks but to their large RH mixing which is not
 observable in  the SM.\\ I would like to present in this talk some
 examples of this ``duality''  but 
before that let me explain what are the RH rotations and why they are important
 and ``observable''.

 {\em What are  RH rotations?}

reiTo  diagonalize   a general complex (mass) matrix M one needs
a bi-unitary transformation, i.e.
two unitary  matrices $U_{L,R}$~, such that
\be
{U_L}^\dagger  M U_R = M_{diagonal}
\ee
or
\be
{U_L}^\dagger MM^\dagger U_L = (M_{diag.})^2 = {U_R}^\dagger M^\dagger
MU_R\quad.
\ee
 
Only in the case of hermitian (symmetric) matrices is $U_R$ related 
to $U_L$
\be
M=M^\dagger (M^T) \ \Longrightarrow \  U_R=U_L({U_L}^*)\quad.
\ee
RH  fermions are singlets in the SM and only LH charged currents
are involved in the weak interactions
\be
{\cal L}_W = W_\mu{\overline {u_L}} \gamma ^\mu V_{CKM} d_L \  + \   h.c.
\ee
where
$$
V_{CKM} = {U_L^u}^\dagger U_L^ d\quad.
$$
The neutral currents are not affected by RH mixing (as long as the 
$U_R's$ are unitary).
However RH rotations are involved in many extensions of the SM,
especially in GUTs. They affect therefore phenomena like:~\footnote{
Note, that the observation of RH contributions to B decay is still an open
possibility~\cite{B}. Also, the $\Lambda_b$ polarization observed in
ALEPH~\cite{A} may be an indication for a RH effect.}\\  
(i) Proton decay,\\
(ii) Neutrino oscillations,\\
(iii) Leptogenesis via decays of RH neutrinos as the origin of baryon 
asymmetry~\cite{bar}, \\
(iv) R-parity violating or leptoquark induced contributions,\\
(v) Radiative corrections etc.

The SM must be extended (at least) to explain the origin of the
fermionic mass matrices, hence:
{\centerline{ \em one cannot simply neglect the RH rotations.}}\\
Large RH mixing angles are not unnatural and can be induced via a
symmetry.
E.g. using  $P_{LR}$ invariance~\cite{plr} -- a generalization of Parity
for gauge theories. One can show that a strong RH rotations of the
light families will force the proton to decay mainly into kaons~\cite{akb}
$$
P \lra  \nmk  \qquad \hbox{and}  \qquad  P \lra \muk .
$$
even without SUSY. But also in SUSY-GUTs the branching ratios of the 
proton decay will be drastically changed~\cite{bpw}~\cite{ar}.

{\em Down-Lepton Duality} 

${\bf SU(5)}$:     The duality comes about naturally already in this minimal 
GUT.\\ This is because
the fermions are distributed in the families $({\bf\bar5} 
+{ \bf10})^i ,\quad i=1,2,3 .$,
in such a  way that $e_L$ and ${d_L}^c$ are always in the same
representation, but not $e_L$ and $d_L$ etc.\\ 
$$
\psi({\bf\bar 5}) = \left(\begin{array}{c}
                          \vec{d^c}\\
                          e^-\\
                          \nu
                      \end{array}\right)_L \qquad
\chi({\bf10}) = (\vec{u^c}, \left( \begin{array}{c}
                                   \vec{u} \\ \vec{d} \end{array}
                                   \right),
                          e^c)_L\quad.         
$$
Hence, if only $H_{\bf\bar{5}}$ and $H_{\bf{5}}$ are used to give the 
fermions masses, as e.g. in
$$
y_{ij}\ \psi_{\bf\bar{5}}^i\ {\bf\chi_{10}}^j <H_{\bf\bar{5}}>
$$
one obtains the mass relation  
$$ 
\large{m^\ell} = (\large{m^d})^T 
$$
 which realizes the {\em d\/} - $\ell$ {\em duality\/}~\cite{5}.
However, $H_{\bf\bar{5}}$ and $H_{\bf{5}}$ alone cannot 
account for the masses of the light families and  adding other Higgs
representations will break this duality. 

${\bf SO(10)}$: The whole family (with $\nu_R$) is here
in one irrep. $\Psi_{\bf{16}}$, 
but there is a ``memory'' of $SU(5)$ in the coupling. It is possible,
therefore, to use combinations of symmetric and antisymmetric couplings
to generate asymmetric mass matrices which give approximately the $d$ - $\ell$
duality. This is usually done in terms of a broken $U(1)$ family
group and the use of non-renormalizable contributions~\cite{10}.
However, the matrix elements are fixed in this framework only  up to
$O(1)$ factors. 
This leads to factors $[O(1)]^3$ in the see-saw neutrino mass
matrix, which means that this method is, in general,  not suitable for
predictions in the neutrino sector.\\
It is possible also to use models without non-renormalizable
contributions and a conserved $U(1)$ or $ Z_n$. This symmetry dictates
the couplings and the mass matrices textures such that 
the experimentally known parameters, i.e. masses and mixing of the
quarks in addition to the masses of the charged leptons, dictate (up to one
parameter) the properties of the neutrinos~\cite{am}. $SO(10)$ is
broken in this model, as follows:
\be
SO(10){}\stackrel{M_U}{\longrightarrow} {}SU_C(4) \times SU_L(2) \times SU_R(2)
\ \stackrel{M_I}{ \longrightarrow} \  SM ,
\ee
where $M_U$ and $M_I$ are fixed by the requirement of unification to be:
\be
\begin{array}{cc}
M_U = 1.31 \times 10^{16}\quad & \quad M_I = 6.14 \times 10^{10} 
\end{array}
\ee
\nopagebreak
in terms of the required Higgs representations.\\
One can choose then the free parameter such that the solution will
account for the observed atmospheric neutrino anomaly together with
either  large-angle MSW or small-angle MSW. Also, specific predictions
for the Proton
(and Neutron) decay branching ratios are obtained, which differ from
the conventional $SO(10)$ prediction~\cite{lang}.
\begin{center}
\begin{tabular}{||l|c|c|c||}
{\em channel}&{\em conventional} &{\em LA MSW}&{\em SA MSW} \\
&  (Kane + Karl)~\cite{kk} &
(Achiman+Merten) & (Achiman+Merten) \\ \hline \hline
$e^+\pi^0$ & 38\% & 21\% & 25\% \\
$\nu\pi $ & 15\% & 35\% & 36\% \\
$\frac{e{^+}\pi^0}{\nu\pi}$&25&0.6&0.7\\
$\mu^+\pi^0$& $\sim 0$&8.5\%&6\%\\
$\mu^+K^0$&18\%&2.6\%&1\%\\
$\nu K^+$&$\sim 0$&3.5\%&2.3\%\\
\hline\hline
\end{tabular}
\end{center}
We see that the neutrino channels are enhanced with respect to the charged
lepton ones. 
\be
\tau(e^+\pi^0) = 10^{34\pm1.7}  yrs,
\ee
i.e. in the range of possible observation by s-kamiokande.  

{\em A new kind of model for the  mass matrices in 
$E_6$ GUT}~\cite{e6}\cite{r}.    

${\bf E_6}$:
Under $E_6 \supset SO(10) \times U(1)$  one family decomposes as 
$$
{\bf 27}_{E_6} =  {({\bf 16} + {\bf 10} + {\bf 1})}_{SO(10)}\quad.
$$
The ``exotic fermions'' $({\bf 10} + {\bf 1})_{SO(10)}$ are denoted as follows:
$$
{\bf{10}}_{\bf{5+\bar 5}} = (\vec{D},E^c,N^c) + (\vec{D}^c,N,E) \qquad ; \qquad
{\bf 1_1}= L 
$$ 
(where $D(E)$ are $d(e)$\/-like quarks(leptons)).
 
{\em The main property of this model}~\cite{ya} {\em is obtained by mixing 
the light fermions with these exotic ones}:\\
One starts with all light mass matrices  $m_u$,$m_d$,$m_\ell$ and $m_{\nu_D}$
diagonal and equal (via a light $<H_{\bf{27}}>$).  The heavy Higgs
representations generate VEVs 
in a different direction. These  give  the exotic fermions heavy masses and 
mix them with the light ones. The  mixing leads then to the
``observed'' mass matrices. In particular,
$m_d$ obtains off-diagonal contributions due to mixing with the $D_i$
but $m_u$ remains diagonal as there are no $u$-like exotic fermions. 
(Rosner~\cite{jr} 
suggests that this is the reason why  $m_{d_i}  < m_{u_i} $).
The special thing about this scenario is that it realizes naturally 
the $d-\ell$ duality. 
This is due to the opposite sign of the charges of the leptons ($e(-1)$) and the d-quarks 
($d(2/3)$).  In the Yukawa coupling \quad $d^cD$ \quad  goes therefore with 
\quad $E^ce$ \quad and \quad $D^cd$ \quad
with  \quad $e^cE$ ~\cite{jr}\cite{bk}.   I.e. the mixing of $(d_i)^c$ 
corresponds to the mixing of $e_i$ etc. \\
Hence:\\
\centerline{\em large RH mixing of $d_i$ $\llra$ large LH
  mixing of $e_i$\quad.}
 
In contrast with $SU(5)$~\cite{5} or $SO(10)$~\cite{10} models the
duality here is independent of the details of the Higgs representations.\\
Let me give now some more details of the model. The most general mass
matrix of  
$ d,d^c,D,D^c$~\cite{jr} ($d(I_L=1/2)$ and $d^c,D,D^c $ all have    $I_L=0$)
$$
M_d = \begin{array}{cc}
&  {\begin{array}{cc}
\hspace{-2mm}d~~~ &  D    
\end{array}}\\ \vspace{2mm}
\begin{array}{c}
d^c \\ D^c 
 \end{array}\!\!\!\!\! &{\left(\begin{array}{cc}
\,\,m~~ &\,\bar M~~ \\
\,\,\bar m~~   &\,M~~ \\
\end{array}\right) }~,
\end{array}  \!\!  ~~~~~
$$
$m$ -- the pure ``light''  $3\times 3$ mass matrix (EW scale)\\
$\bar m$ -- the ``light'' $3\times 3$ mixed matrix   ($\Delta I_L=1/2$\ , EW scale)\\
$M$ -- the pure ```heavy''   $3\times 3$  mass matrix ($\Delta I_L=0$\ , heavy 
exotic scale)\\
$\bar M$ -- the mixed ``heavy''  $3\times 3$  mass matrix ($\Delta I_L=0$\ , heavy 
exotic scale).\\
For the charged leptons we have (note that there are here different Clebsch-Gordan factors):\\
$$
M_e = \begin{array}{cc}
&  {\begin{array}{cc}
\hspace{-2mm}e~~~ & E   
\end{array}}\\ \vspace{2mm}
\begin{array}{c}
e^c \\ E^c 
 \end{array}\!\!\!\!\! &{\left(\begin{array}{cc}
\,\,m_1~~ &\,\bar m_1~~ \\
\,\,\bar M_1~~   &\,M_1~~ \\
\end{array}\right) }~,
\end{array}  \!\!  ~~~~~
$$

 To find  the physical states, one must diagonalize  the mass
 matrices. To get the LH rotations one uses :
\be
M_d^\dagger M_d = 
\left (
\begin{array}{cc}
m^\dagger m + \bar m^\dagger \bar m&m^\dagger \bar M + \bar m^\dagger M \\
m^\dagger \bar M + \bar m^\dagger M&\bar m^\dagger \bar M + M^\dagger M \\
\end{array}\right)
\ee
and the RH rotations are obtained by diagonalization of:
\be
M_d M_d^\dagger = 
\left (
\begin{array}{cc}\
m m^\dagger + \bar M \bar M^\dagger &  m \bar m^\dagger+ \bar M M^\dagger \\  
m \bar m^\dagger + \bar M M^\dagger   &  \bar m \bar m^\dagger + M M^\dagger \\\end{array}\right)
\ee
If \quad  $m,\bar m  << M,\bar M$ , \quad $M_d^\dagger M_d$\quad is similar to the see-saw matrix, so that
$$
d \approx d_0 - \frac {\bar m^2}{M^2} D_0 \qquad
D \approx \frac {\bar m^2}{M^2} d_0 + D_0 ,
$$
i.e very small heavy mixing.\\
For $M_d M_d^\dagger$ it is possible to use the same approximation
only if \quad $m,\bar m  << \bar M << M$.\\ 
In this case one obtains for the RH part:
$$
d^c \approx d_0 ^c- \frac {\bar M^2}{M^2} D_0^c\qquad
D \approx \frac {\bar M^2}{M^2} d_0^c + D_0^c ,
$$
i.e. the heavy-light RH mixing is in general larger.
 
The  $3 \times 3$ light-light mixing matrices are in general not
diagonal even if 
$m$ was diagonal.  They are also not unitary but for the LH mixing the deviation is 
very small $O(\frac{m^2}{M^2})$. For the RH rotations however the deviations 
from  unitarity are of $O(\frac{\bar M^2}{M^2})$. Those must be also small to avoid
 inconsistency with experimental limits on the neutral current and in particular 
from \quad $e^+e^-\longrightarrow Z  \longrightarrow b \bar b$.~\footnote{This is an 
indication that the scale of $M$ should  be quite high.  Also, note 
that  already $\frac{\bar M}{M}= \frac{1}{10} $ gives $g_{bR} \delta
g_{bR}\approx 10^{-3}$, hence the heavy scale difference must not be
very large.}
Very heavy $M$ and/or small heavy-light mixing does not exclude
large RH rotations. 
The  Family  mixing in the light-light matrix is more dependent on
the light mixed matrix $\bar m$.\\
One can show~\cite{ya} that it is possible to choose the heavy mass
matrices in such a way that the known masses and mixing are obtained
(at least for two families).

{\em Conclusions about RH rotations}
 
$d-\ell$  duality, especially in $E_6$, is a natural explanation for the 
contradiction between the large LH mixing angles of the leptons and the small
ones of the quarks.
 
Many models are known to be able to give the fermionic masses, 
$V_{CKM}$, \ $\nu$- properties etc. (within the experimental
errors).  This is only an 
indication that the mass question is far from from being  solved. Part of the 
problem is related to the fact that those models disregard the RH rotations.\\
The hope is that correlations between neutrino physics, proton decay, leptogenesis, 
baryon assymetry etc. will tell us something about the RH rotations. This will 
limit considerably the freedom in the fermionic mass matrices.
 
Many recent models use asymmetric mass matrices which induce large RH mixing. 
One cannot simply neglect this fact and must consider possible
implication of the RH rotations to have a complete model.

\end{document}